\documentclass[a4paper]{jpconf}
\usepackage{graphicx,amsbsy}
\begin{document}
\title{Half-magnetization plateaux in Cr spinels}

\author{Nic Shannon$^1$, 
Hiroaki Ueda$^2$,
Yukitoshi Motome$^{3}$,
Karlo Penc$^4$, 
Hiroyuki Shiba$^5$,
and Hidenori Takagi$^{6,7,8}$
}

\address{$^1$H. H. Wills Physics Lab, Tyndall Av., Bristol BS8~1TL, UK}
\address{$^2$ISSP, University of Tokyo, Kashiwa, 277-8581, Japan}
\address{$^3$Department of Applied Physics, University of Tokyo, Bunkyo-ku, Tokyo 113-8656, Japan }
\address{$^4$Research Institute for Solid State Physics and Optics, H-1525 Budapest, P.O.B. 49, Hungary. }
\address{$^5$The Institute of Pure and Applied Physics, 2-31-22 Yushima, Bunkyo-ku, Tokyo 113-0034, Japan}
\address{$^6$Department of Advanced Materials Science, University of Tokyo, Kashiwa, 277-8651, Japan}
\address{$^7$RIKEN(The Institute of Physical and Chemical Research), Wako, 351-0198, Japan}
\address{$^8$CREST, Japan Science and Technology Agency (JST), Kawaguchi, 332-0012, Japan}

\ead{nic.shannon@bristol.ac.uk}

\begin{abstract}
    Magnetization plateaux, visible as anomalies in magnetic susceptibility at low temperatures, are one of the 
    hallmarks of frustrated magnetism. An extremely robust half-magnetization plateau is observed in the spinel oxides 
    CdCr$_2$O$_4$ and HgCr$_2$O$_4$, where it is accompanied by a substantial lattice distortion. We give an overview of the present state 
    experiment for CdCr$_2$O$_4$ and HgCr$_2$O$_4$, and show how such a half-magnetization plateau arises quite naturally 
    in a simple model of these systems, once coupling to the lattice is taken into account.
\end{abstract}

\section{Introduction}

Spinels AB$_2$X$_4$, where A and B are metal ions and X=\{O, S, Se\ldots\}, are among the most ubiquitous of crystal structures.   
While the undistorted spinel lattice has overall cubic symmetry, the network of B--ion sites forms a highly frustrated pyrochlore lattice, built 
of corner--sharing tetrahedra.   Simple nearest--neighbour interactions 
on the pyrochlore lattice cannot select a unique magnetic 
or charge ordered ground state~\cite{anderson56}.  
It is this fact which makes the properties of magnetic spinels --- and 
in particular their high field behaviour --- so interesting.

\begin{figure}[h]
\begin{minipage}{24pc}
\includegraphics[width=24pc]{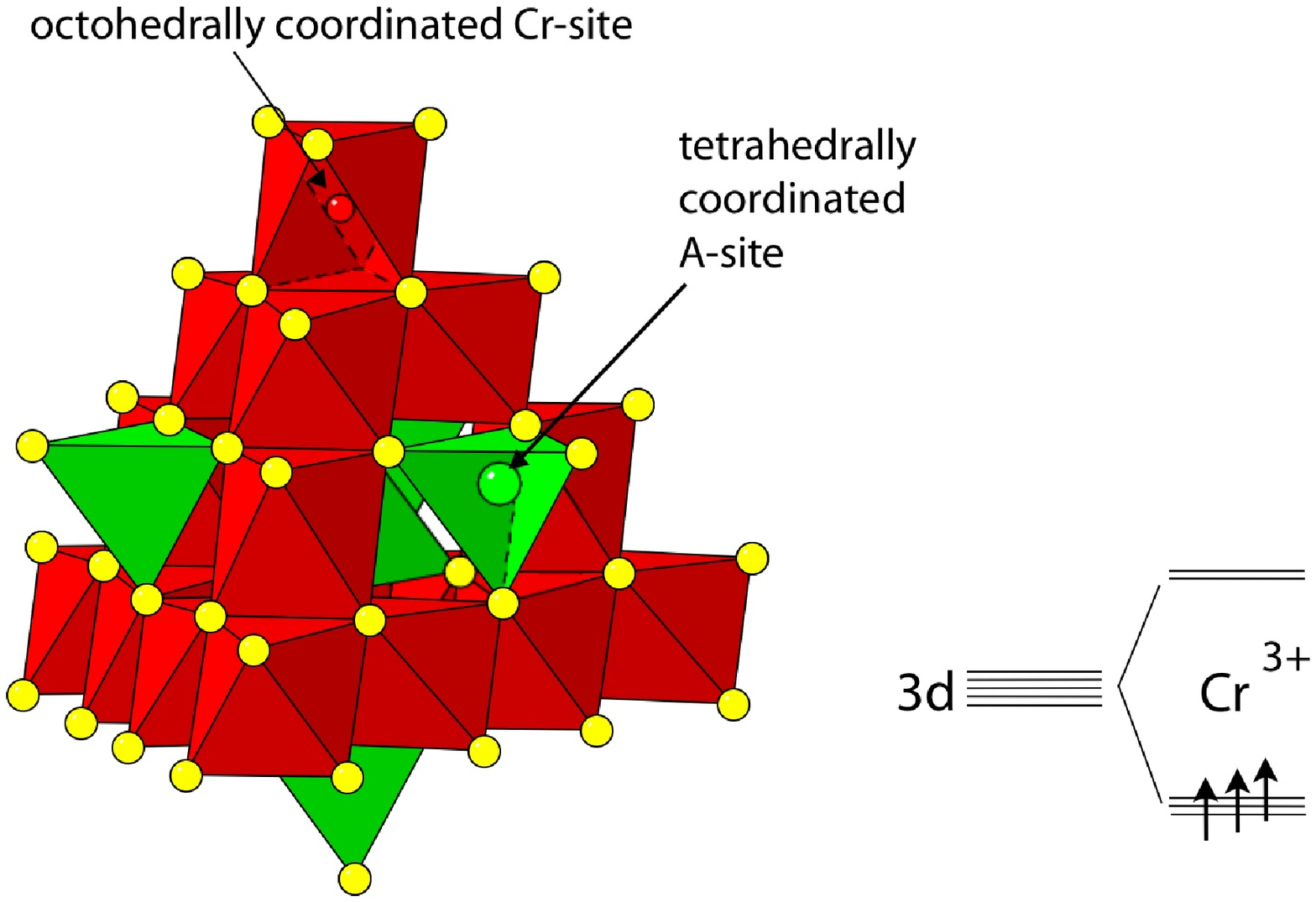}
\caption{\label{fig:spinel}Structure of spinel oxide ACr$_2$O$_4$.  O$^{2-}$ ions 
shown in yellow define an octahedral environment for the Cr$^{3+}$ 
ions, which have a full $t_{2g}$ shell of $3d$ electrons with moment $S=3/2$.
Detailed control of lattice parameters can be achieved by substitution of different A-site metal ions.}
\end{minipage}\hspace{2pc}%
\begin{minipage}{12pc}
\includegraphics[width=12pc]{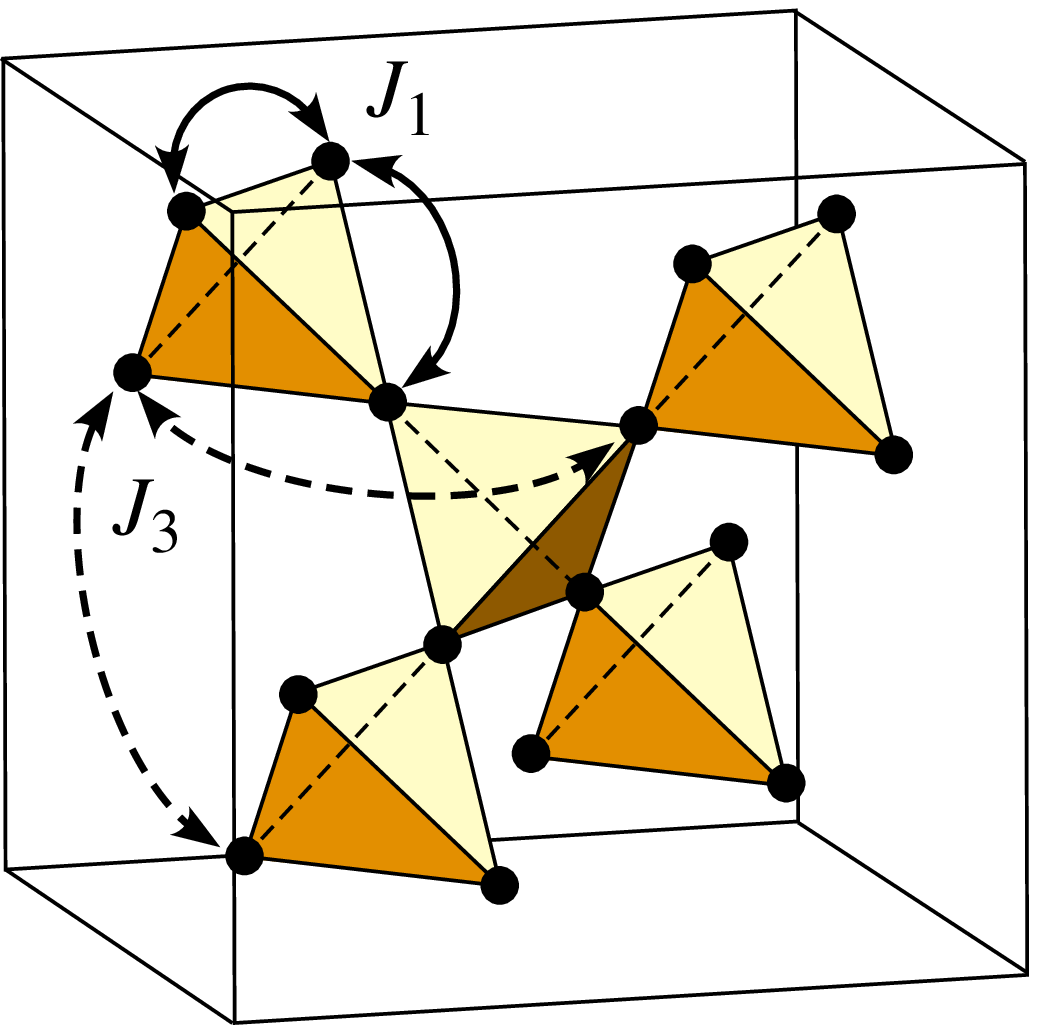}
\caption{\label{fig:pyrochlore}Cubic unit cell of pyrochlore lattice, 
showing first-- and third--neighbour bonds.
}
\end{minipage} 
\end{figure}

Chromium spinel oxides ACr$_2$O$_4$ offer the opportunity to study magnetic 
frustration in the absence of charge and orbital effects.
The Cr ion has a strong Hund's rule coupling and lives in an octahedral crystal field. 
Therefore, if A=\{Zn, Cd, Hg\ldots\} 
is a divalent metal ion, the Cr will be in a $[Ar]3d^3$ high--spin 
state for which all of its $t_{2g}$ symmetry orbitals are singly occupied, and all of the $e_{g}$ symmetry 
orbitals are empty --- see Figure~\ref{fig:spinel}.   The resulting 
bulk state is a Mott insulator in which all charge and orbital effects are 
quenched.  The magnetic properties of a spinel oxide like 
CdCr$_2$O$_4$ therefore depend by  the way in which a set of 
$S=3/2$ spins behave on a pyrochlore lattice.

This is a far from simple problem.  
The minimal magnetic model suggested by the chemical structure of ACr$_2$O$_4$ is 
\begin{eqnarray}
{\mathcal H}_0 &=& J_1 \sum_{\langle ij\rangle_1}  {\bf S}_i {\bf 
S}_j - {\bf h} \sum_i {\bf S}_i
\label{eqn:H0}
\end{eqnarray}
where $J_1$ is a Heisenberg exchange interaction, ${\bf h}$ the 
applied magnetic field and the sum $\langle ij\rangle_1$ runs over the nearest--neighbour 
bonds of a pyrochlore lattice (Figure~\ref{fig:pyrochlore}).  
In conventional magnetic insulators, exchange interactions select a unique 
ground state for which the classical energy per bond is minimized.
However, because of the special corner--sharing geometry of the pyrochlore lattice, 
the classical energy of any state can be rewritten as
\begin{eqnarray}
    E_0 = 4 J_1\sum_{\mbox{tetr.}} 
     \left({\bf M} - \frac{{\bf h}}{8J_1}\right)^2 -\frac{h^2}{16J_1} + \mbox{const.} 
\label{eqn:E0}
\end{eqnarray}
where \protect\mbox{${\bf M} = ({\bf S}_1 + {\bf S}_2 + {\bf S}_3 +  {\bf S}_4)/4$} 
is the magnetization per spin of a given tetrahedron and the sum runs 
over all tetrahedra.     
The three components of ${\bf M} = (M_x, M_y, M_z)$ fix only three of the four 
independent classical spin angles per tetrahedron.  
Thus nearest--neighbour interactions do not select 
a unique ground state, but rather an entire manifold of states for 
which ${\bf M} \equiv {\bf h}/8J_1$ in every tetrahedron.
Cr spinels can therefore be expected to exhibit very exotic magnetic properties.

\begin{center}
\begin{table}[h]
\caption{\label{table1} Ordering temperatures and crystal symmetries of Cr spinels in zero 
applied field.  Values of $T_N$ and $\theta_{CW}$ are for samples 
prepared by Ueda et al.~\cite{ueda06}.   
The nearest neighbour exchange $J_1$ is estimated from $\theta_{CW} \approx z J_1 S(S+1)/3$.}
\centering
\begin{tabular}{@{}*{7}{l}}
\br
 & MgCr$_2$O$_4$~\cite{rovers02} & ZnCr$_2$O$_4$~\cite{lee00} & 
 CdCr$_2$O$_4$~\cite{rovers02} & HgCr$_2$O$_4$~\cite{ueda06} \\
\mr
$T_N$ [K] & 12.5 & 12 & 8 & 5.8 \\
$\theta_{CW}$ [K] & 370 & 390 & 70 & 32 \\
$T_N/\theta_{CW}$ & 0.03 & 0.03 & 0.11 & 0.18\\
\mr
$J_1$ [K] &  49  & 53 & 9 & 4 \\
\mr
Unit cell ($T < T_N$) & tetragonal, $c <a$ 
   & tetragonal, $c <a$ & tetragonal, $c > a$ & 
   orthorhombic\\
\br
\end{tabular}
\end{table}
\end{center}

At high temperatures, Cr spinels exhibit very little anisotropy in their magnetism.   
Their susceptibility has the form $\chi = C/(T + \theta_{CW})$, where 
\mbox{$C \approx 1.9$ emu/mol K} is  
in the range expected for a $S=3/2$ moment with $g_{L} \approx 2.0$.
Values of $\theta_{CW}$ vary from $370$~K in MgCr$_2$O$_4$ to $32$~K 
in HgCr$_2$O$_4$ (see Table~\ref{table1}), suggesting dominantly
antiferromagnetic (AF) exchange interactions between Cr spins.
However the Curie--Weiss form of the magnetic susceptibility persists down 
to very low temperatures, and magnetic order is achieved only for 
$T < T_N \ll \theta_{CW}$, where it is accompanied by a structural transition.   
In CdCr$_2$O$_4$, ZnCr$_2$O$_4$ and MgCr$_2$O$_4$, this changes the 
overall crystal structure from cubic to tetragonal and introduces a 
weak easy--plane anisotropy in the magnetism; in HgCr$_2$O$_4$
the change is from cubic to orthorhombic (Table~\ref{table1}).

The thermodynamic properties of ACr$_2$O$_4$ at high temperature and 
in zero field are typical of a wide range of frustrated 
magnets, and entirely compatible with the minimal model Equation~\ref{eqn:H0}.
The special feature of these systems is the structural 
transition accompanying magnetic order at low temperature.   
This enables the system to lift the infinite classical ground state 
degeneracy  exhibited by Equation~(\ref{eqn:E0})--- a mechanism dubbed 
{\it order by  distortion}~\cite{tchernyshyov02}.

The question addressed in this paper is what happens to the magnetic 
properties of Cr spinels in applied magnetic field.
The scale of exchange interactions in the Cd and Hg compounds
implies the bulk of their magnetization process occurs over
a range of fields now routinely accessible in large pulsed magnets.
As we shall see, the {\it order by distortion} mechanism remains active in applied 
magnetic field, and has some very dramatic consequences.

\section{Magnetization process and half--magnetization plateau}

The magnetization process $M(h)$ of a classical Heisenberg AF on 
the pyrochlore lattice at zero temperature is linear up to a 
saturation field of $8J_1$ --- c.f. Equation~(\ref{eqn:E0}).
Under the assumption that longer range interactions are weak, 
values of $J_1 \approx 9$~K and $J_1 \approx 4$~K 
can be inferred from $\theta_{CW} \approx z J_1 S(S+1)/3$ 
for CdCr$_2$O$_4$ and HgCr$_2$O$_4$, respectively (Table~\ref{table1}).
These  suggest that the saturation field is experimentally 
accessible for both compounds.

\begin{figure}[h]
\begin{minipage}{18pc}
\includegraphics[width=18pc]{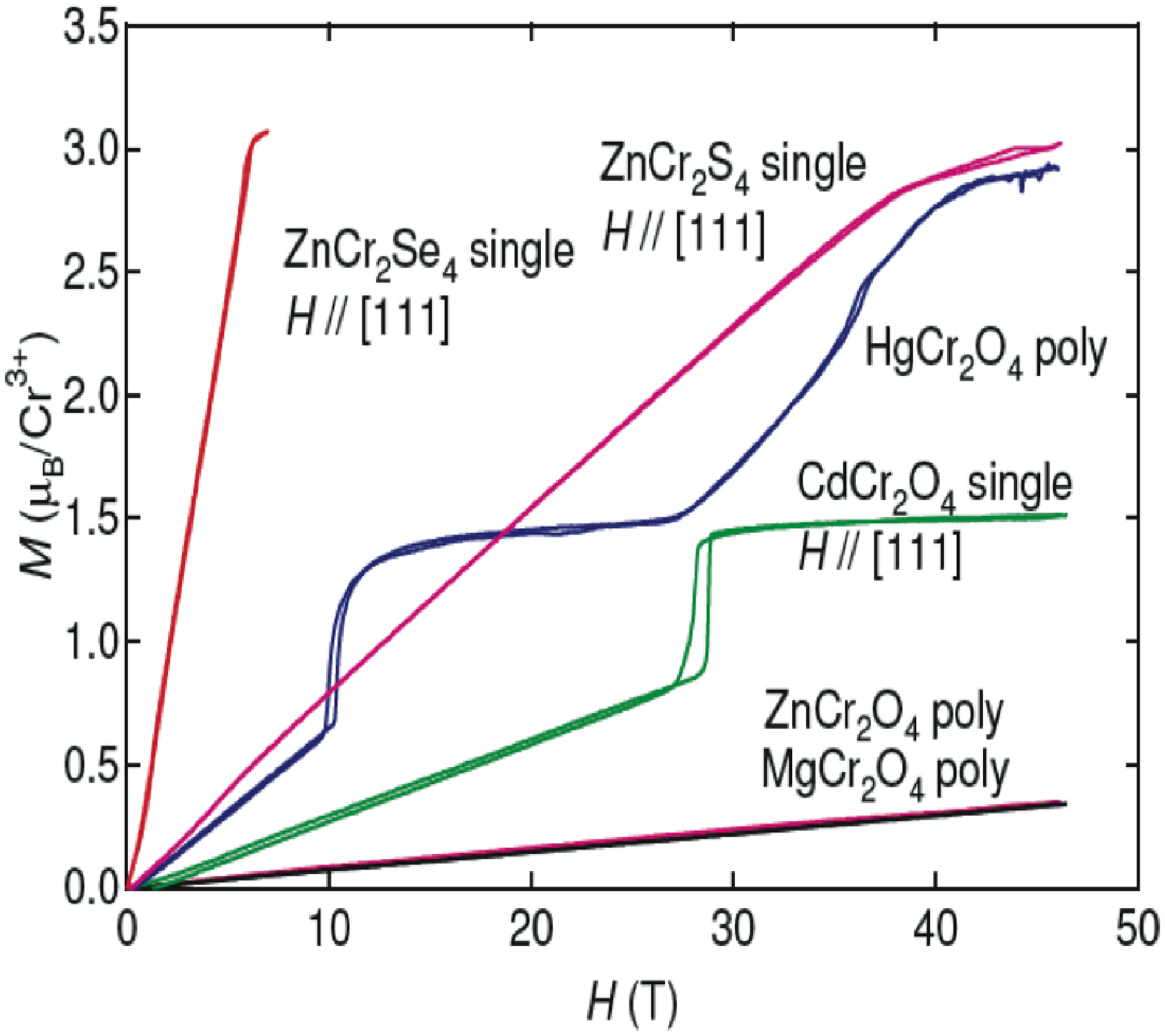}
\caption{\label{fig:plateau}Magnetization processes of a variety of 
single crystal and polycrystalline samples of Cr spinels taken 
at 1.8K, showing dramatic half--magnetization plateaux in CdCr$_2$O$_4$ and HgCr$_2$O$_4$.
Data taken from~\cite{ueda06}.}
\end{minipage}\hspace{2pc}%
\begin{minipage}{18pc}
\includegraphics[width=18pc]{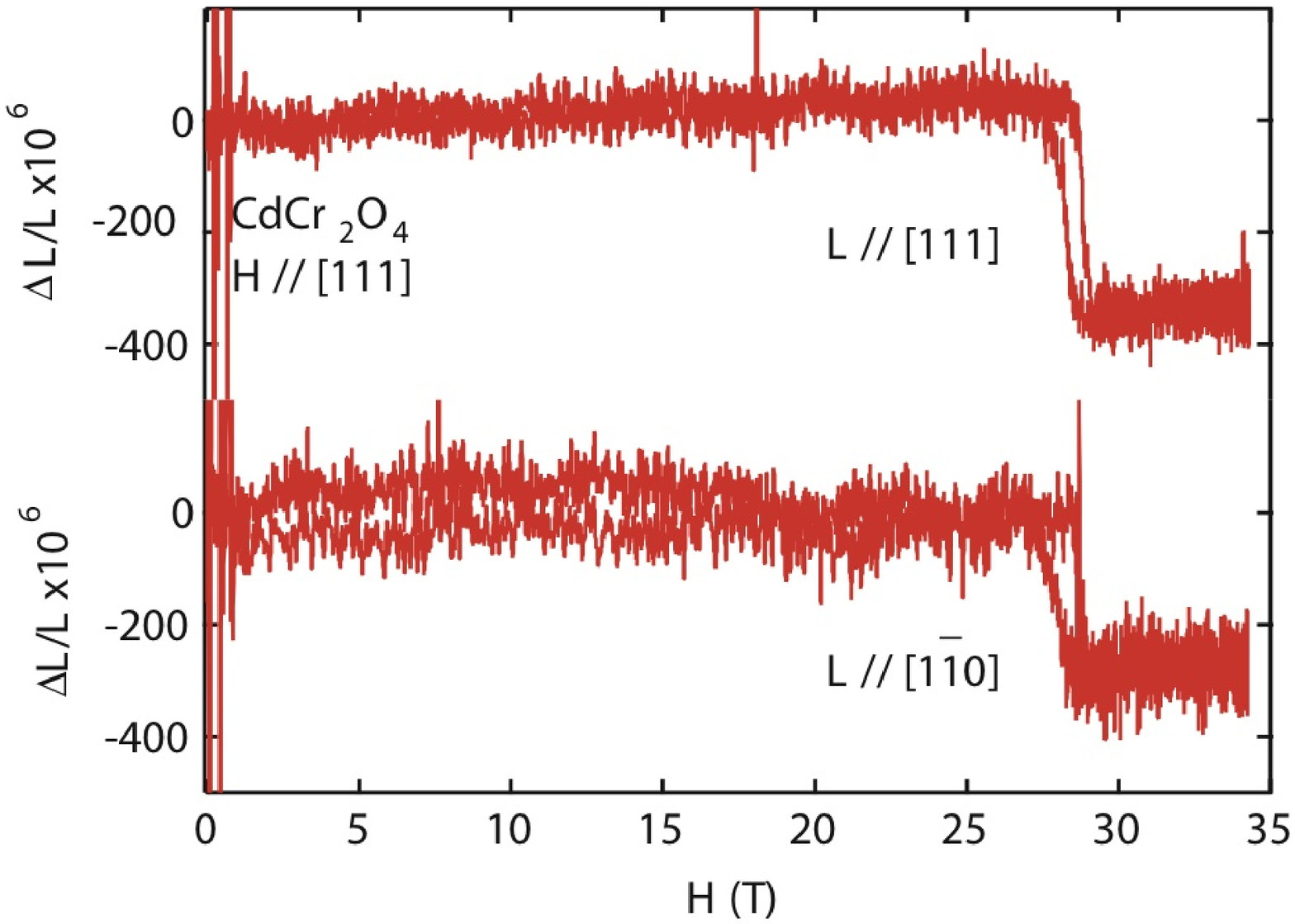}
\caption{\label{fig:magnetostriction}Colossal {\it negative} magnetostriction 
accompanying first--order transition into half--magnetization plateau 
state in CdCr$_2$O$_4$.  Data taken from~\cite{ueda05}.}
\end{minipage} 
\end{figure}

Measurements of $M(h)$ for CdCr$_2$O$_4$ and ZnCr$_2$O$_4$
were initially made in pulsed magnetic fields of up to $48$~T at the 
ultrahigh magnetic field facility of the Institute for Solid State 
Physics (ISSP) in Tokyo~\cite{ueda05}.   
More recently these measurements have been 
repeated for a range of Cr spinels, including 
HgCr$_2$O$_4$~\cite{ueda06}.  Measurements on CdCr$_2$O$_4$ 
have been made at fields of up to $70$~T at the High Magnetic Field Laboratory in 
Osaka~\cite{hagiwara-unpub}, and preliminary measurements on the same 
system at fields of up to $100$~T made at ISSP~\cite{mitamura05}.

Results for a variety of Cr spinels taken at ISSP are shown in Figure~\ref{fig:plateau}.
The magnetization of all compounds is indeed linear over a wide range 
of (low) magnetic fields.   However both CdCr$_2$O$_4$ and HgCr$_2$O$_4$
exhibit a broad magnetization plateau for intermediate values of 
magnetic field.  
The plateau in CdCr$_2$O$_4$ has its onset at $28$~T~\cite{ueda05}, and extends
up to approximately $60$~T~\cite{mitamura05,hagiwara-unpub}.
It has a magnetization of 1.5$\mu_B$ per Cr --- exactly 
half the full moment of an $S=3/2$ Cr$^{3+}$ ion.   
The plateau in HgCr$_2$O$_4$ extends from $10$~T to $27$~T, with a 
saturation field approaching $50$~T.   Once again the plateau occurs 
for exactly half the saturation magnetization.
It is somewhat more rounded than in CdCr$_2$O$_4$; this can 
be attributed to disorder in the polycrystalline sample used~\cite{ueda06}. 

The transition into the plateau state is strongly first--order in both 
cases, displaying a marked hysteresis, and is accompanied by a
large change in the unit cell volume of \mbox{$\mid \Delta V \mid /V \approx 0.33 \times 10^{-3}$ in CdCr$_2$O$_4$}
and \mbox{$\mid \Delta V \mid /V \approx 1.6 \times 10^{-3}$} in HgCr$_2$O$_4$~\cite{ueda06}.
In CdCr$_2$O$_4$, where good single crystals are available, a colossal 
{\it negative} magnetostriction $\Delta L/L = -4 \times 10^{-4}$ can be resolved 
parallel to field in the $[111]$ direction, accompanied by a 
magnetostriction of similar order in the $[1\overline{1}0]$ direction 
(i.e. perpendicular to the field) --- see Figure~\ref{fig:magnetostriction}.   
This scale of magnetostriction would not be unusual in a system with 
Jahn--Teller active $e_g$ orbitals, but is exceptional for an oxide 
in which orbital degrees of freedom are quenched. 
Taken together these, results suggest that the magnetization plateau 
has its origin in a very strong exchange striction.  Coupling to the 
lattice appears to be somewhat stronger in HgCr$_2$O$_4$ than 
in CdCr$_2$O$_4$.   This is also reflected in a higher ratio 
of $T_N/\theta_{CW}$ for the magnetically ordered phase in zero field 
(Table~\ref{table1}). 

The transition out of the plateau state at high field is of second order in both compounds.
Above this second critical field, the magnetization process is again 
approximately linear, suggesting that the new phase arises simply 
through a canting of the ordered moments of the plateau phase. 
For HgCr$_2$O$_4$, it is possible to map out
the entire magnetization process up to saturation at $h \approx 45$~T --- see Figure~\ref{fig:plateau}.  
A further, first--order magnetic phase transition is observed at $h \approx 
37$~T.
While no direct structural data measurements presently exist, this is presumably 
also accompanied by a change in crystal structure or bond length.

\begin{figure}[h]
    \begin{minipage}{18pc}
    \includegraphics[width=18pc]{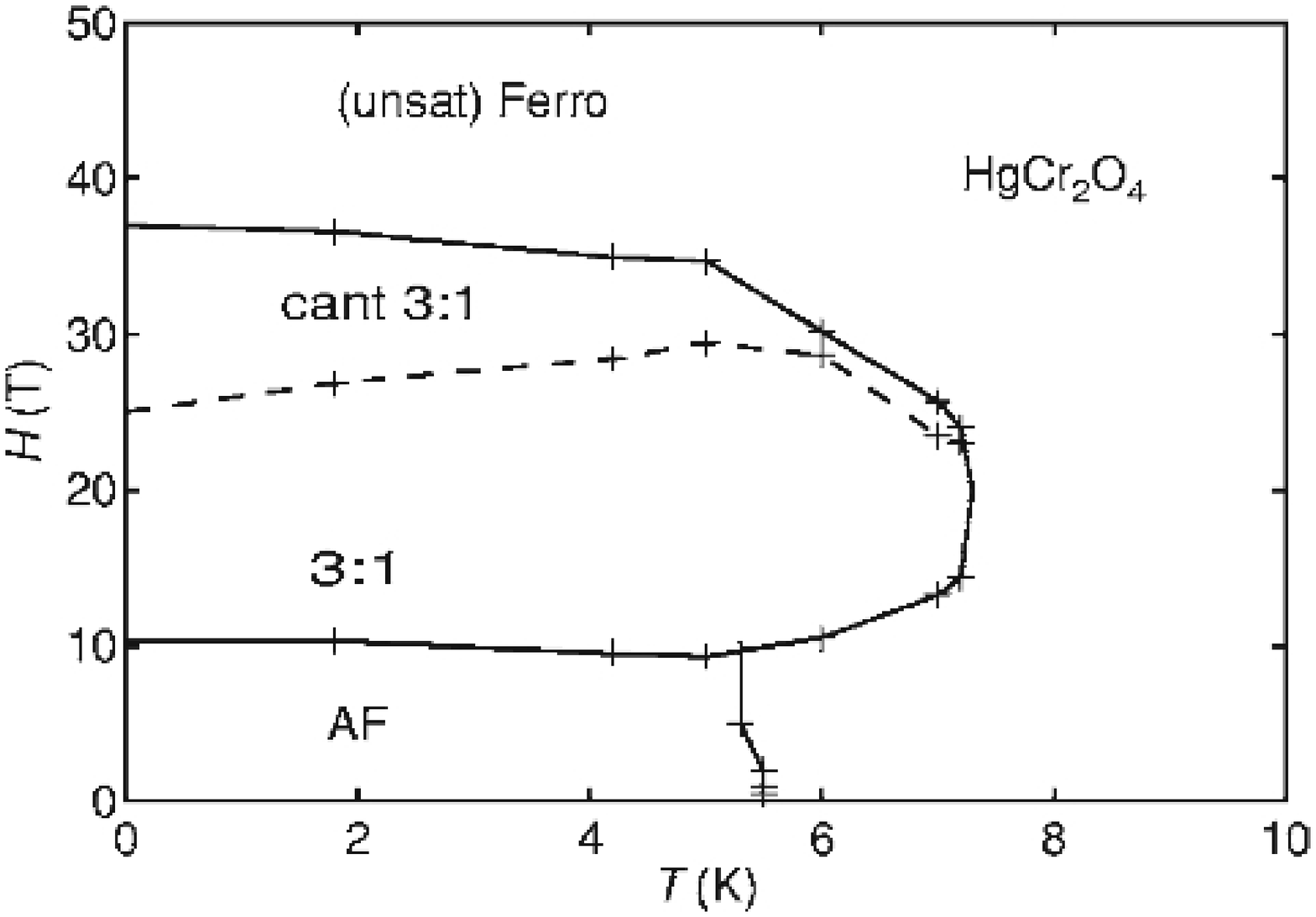}
    \caption{\label{fig:finiteT-experiment}Magnetic phase diagram of
    HgCr$_2$O$_4$ as a function of temperature, taken 
    from~\cite{ueda06}.  Solid lines denote first order phase 
    transitions; dashed lines second order.}
    \end{minipage} 
    \hspace{2pc}%
\begin{minipage}{18pc}
\includegraphics[width=18pc]{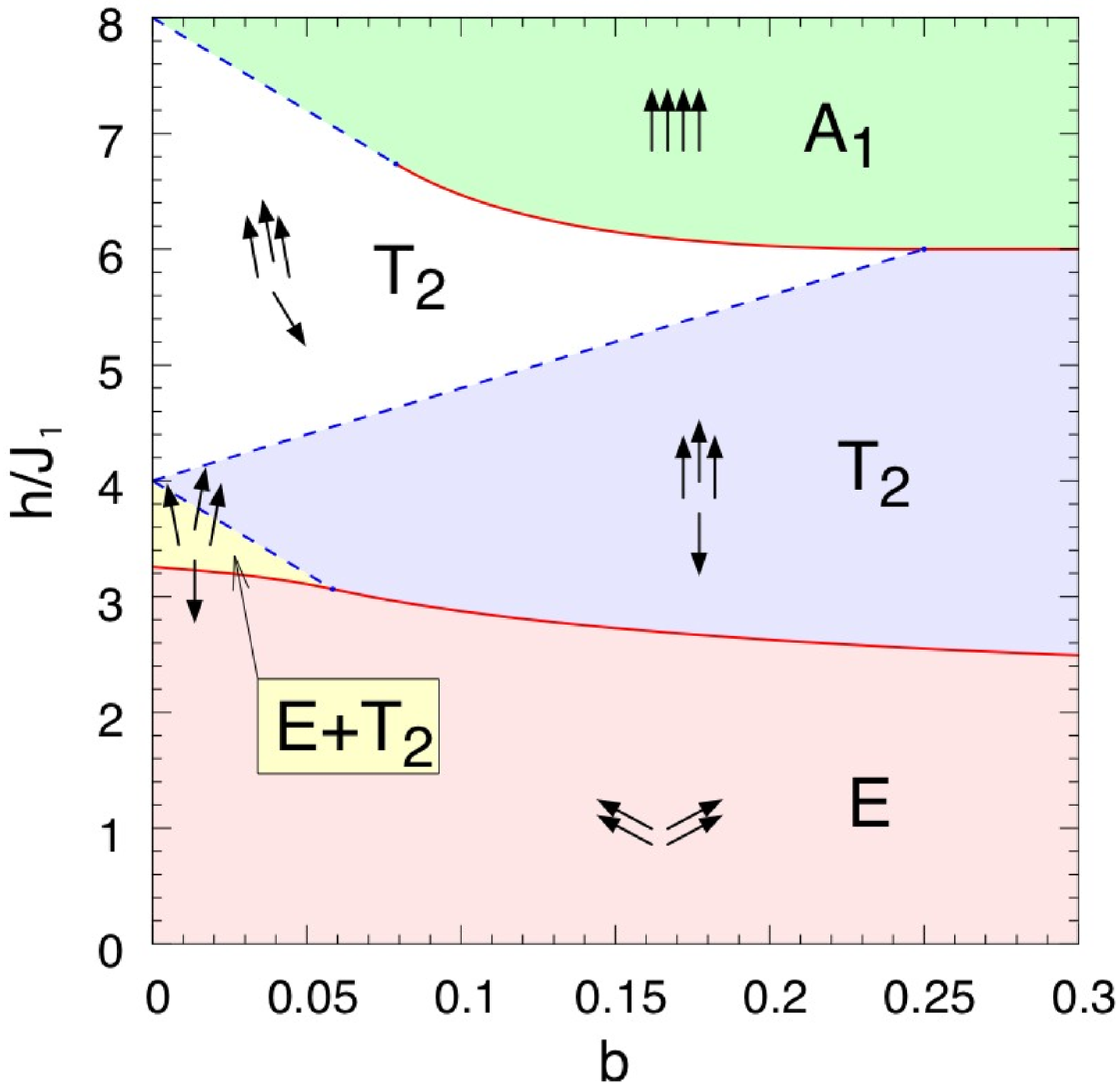}
\caption{\label{fig:theory2}Phase diagram of effective spin 
model Equation~(\ref{eqn:Hb}) taken from~\cite{penc04}, as a 
function of magnetic field $h$ and dimensionless coupling constant $b$. 
Solid lines denote first order phase transitions; dashed lines 
second order. 
All order parameters can be classified in terms of the irreps of the tetrahedral 
symmetry group $T_d$.}
\end{minipage} 
\end{figure}

As the temperature of the sample is raised, the magnetization plateau 
acquires a finite slope.  Above a characteristic transition temperature 
$T_c \approx 10$~K (CdCr$_2$O$_4$), $T_c \approx 7$~K (HgCr$_2$O$_4$)
the plateau washes out entirely and the magnetization process $M(h)$
remains linear up to fields approaching saturation.   
Thermal fluctuations clearly contribute to the 
stability of the plateau state; its width in $h$ increases with increasing 
temperature $T$, and its transition temperature 
is somewhat higher than that for the (canted) N\'eel order at low fields.
Quantum and thermal fluctuations are known to favour collinear spin 
configurations~\cite{henley89}, and in the light of these results it seems natural to 
attribute the half--magnetization plateau to a collinear ``$uuud$'' spin 
configuration with three up and one down spins per tetrahedron.
It is also interesting to note that the critical field for the transition 
from the canted N\'eel state into the half--magnetization plateau is 
almost independent of temperature.

In summary --- the magnetic 
phase diagram for HgCr$_2$O$_4$ deduced from pulsed field measurements
is shown in Figure~\ref{fig:finiteT-experiment}.    It exhibits four 
phases, which we can identify as :
\begin{itemize}
\item a canted N\'eel state at low field 
\item a collinear half--magnetization plateau state for intermediate field 
\item a second canted state approaching saturation
\item a saturated paramagnetic state for large magnetic field
\end{itemize}
Recent elastic neutron scattering data suggest that the half--magnetization plateau 
state in HgCr$_2$O$_4$ has cubic symmetry and a 16--sublattice magnetic 
order~\cite{matsuda--unpub}.  The saturated paramagnetic state at high field is, 
presumably, also cubic.  We return to these points below.

\section{Theory of half--magnetization plateau }

Magnetization plateaux are very common in frustrated magnets.   They 
are known to occur, for example, in both the Ising~\cite{miyashita86} and 
Heisenberg~\cite{kawamura85,chubukov91} models on the triangular lattice.
The conventional explanation for magnetization plateaux is that 
quantum and/or thermal fluctuations act to favour a collinear state 
in which all spins point either parallel or antiparallel to the 
applied magnetic field.   This state may (or may not) exhibit long 
range magnetic order, but since the number of spins in the system is
an integer, it must exhibit a magnetization which is a rational 
fraction of the full moment --- $M=1/3$ in the case of the triangular lattice.
This leads to plateaux which are weak (being a fluctuation 
effect), and uncorrelated with lattice distortion.

The essential features of the half--magnetization plateau seen in Cr 
spinels are its extreme robustness and the large accompanying lattice distortion.   
Any model which seeks to explain it should therefore include the 
coupling between spins and the lattice, as well as the interactions 
between spins.   
We therefore consider the extended Heisenberg model
\begin{eqnarray}
 \mathcal{H} &=&  \sum_{\langle ij \rangle_1} 
  \left[
  J_1 (1- \alpha_1 \rho_{ij})
  {\bf S}_i {\bf S}_j
 + \frac{K}{2}  \rho_{ij}^2 \right] 
- {\bf h} \sum_{i} {\bf S}_i \;,
\label{eqn:H}
\end{eqnarray}
where $K$ is an elastic constant, and the bond variable $\rho_{ij}$ measures the change in 
separation of two Cr ions relative to the state with the lowest elastic energy~\cite{penc04}.

The physics of the system is most easily demonstrated if we make two further simplifying 
assumptions ---  i) we treat the spins as classical variables ${\bf 
S}_i = (\sin\theta_i \cos\phi_i , \sin\theta_i \sin\phi_i, \cos\theta_i )$ and ii) we 
assume that the spins couple to phonons with crystal momentum 
${\bf q} = 0$, i.e. that there is an overall 4--sublattice order.
Under these assumptions, we can eliminate the lattice variables 
$\rho_{ij}$ from the problem entirely and work with the pure spin model
\begin{equation}
 \mathcal{H} =  J_1 \sum_{\langle ij \rangle_1} 
\left[
   {\bf S}_i {\bf S}_j
   - b ({\bf S}_i {\bf S}_j)^2 \right] -  {\bf h} \sum_{i} {\bf S}_i
    \;.
   \label{eqn:Hb}
\end{equation}
where $b= J \alpha^2/2 K$ is a dimensionless measure of spin--lattice coupling.  
The biquadratic term in Equation~(\ref{eqn:Hb}) is of a form
which favours coplanar or collinear spin configurations, and as such is known 
to mimic quantum and thermal fluctuations \cite{henley89}.   
A more general choice of spin--lattice Hamiltonian can lead to more 
complex interactions in the effective spin model.   
However these do not change the essential physics of the problem, and 
for our purposes it will generally be sufficient to enforce  four--sublattice 
order by adding a ferromagnetic (FM) third--neighbour interaction $J_3 < 0$
\begin{eqnarray}
\Delta {\mathcal H}_3 = J_3 \sum_{\langle ij \rangle_3}  {\bf S}_i {\bf S}_j
\end{eqnarray}
to the effective spin model Equation~\ref{eqn:Hb} (see Figure~\ref{fig:pyrochlore}).  
The consequences of relaxing the assumptions (i) and (ii) will be discussed below.

\begin{figure}[h]
    \begin{minipage}{18pc}
    \includegraphics[width=18pc]{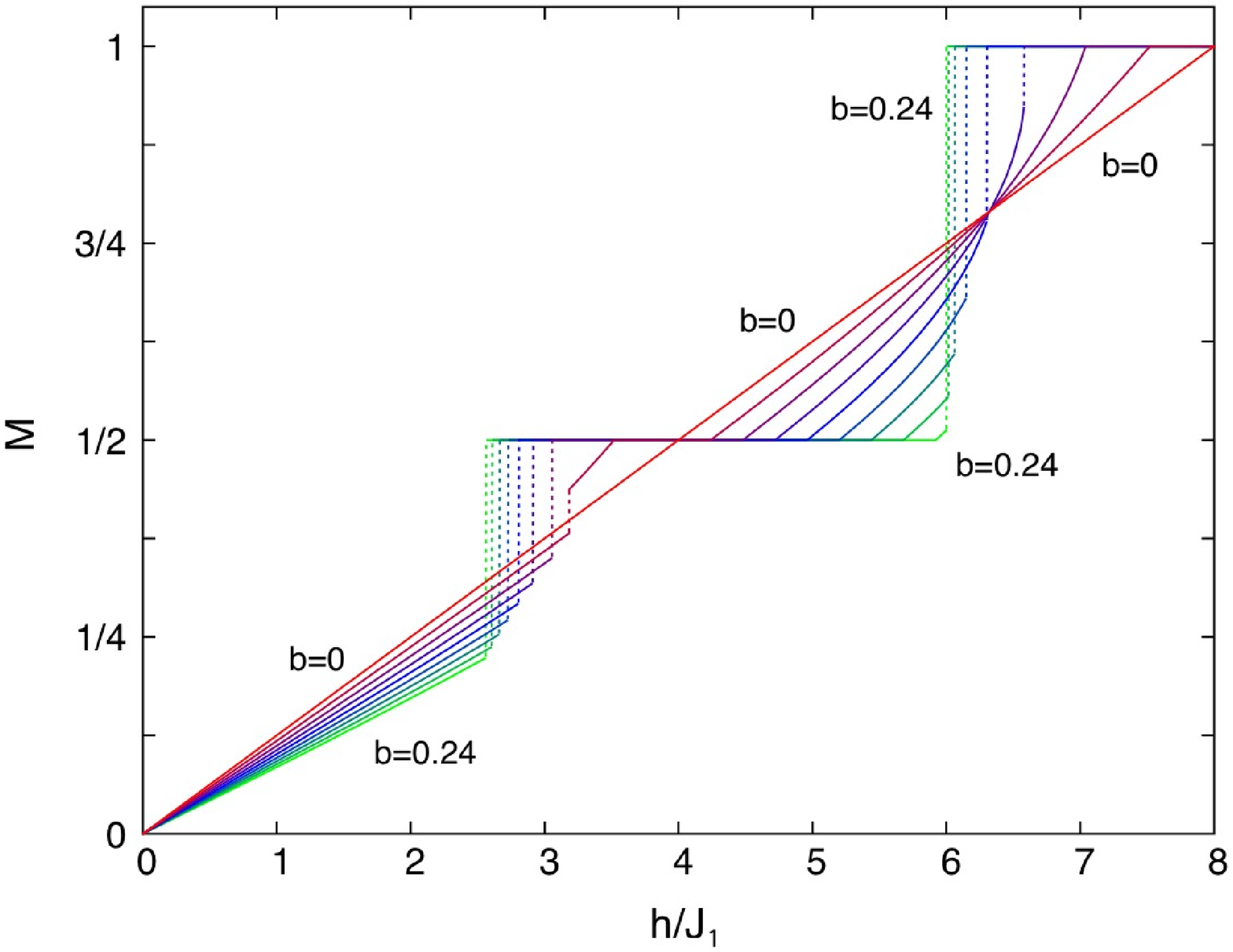}
    \caption{\label{fig:theory1}Magnetization process of effective spin 
    model Equation~(\ref{eqn:Hb}) taken from~\cite{penc04}, 
    for a range of values of the dimensionless spin--lattice coupling $b$.}
    \end{minipage}
    \hspace{2pc}%
    \begin{minipage}{18pc}
\includegraphics[width=18pc]{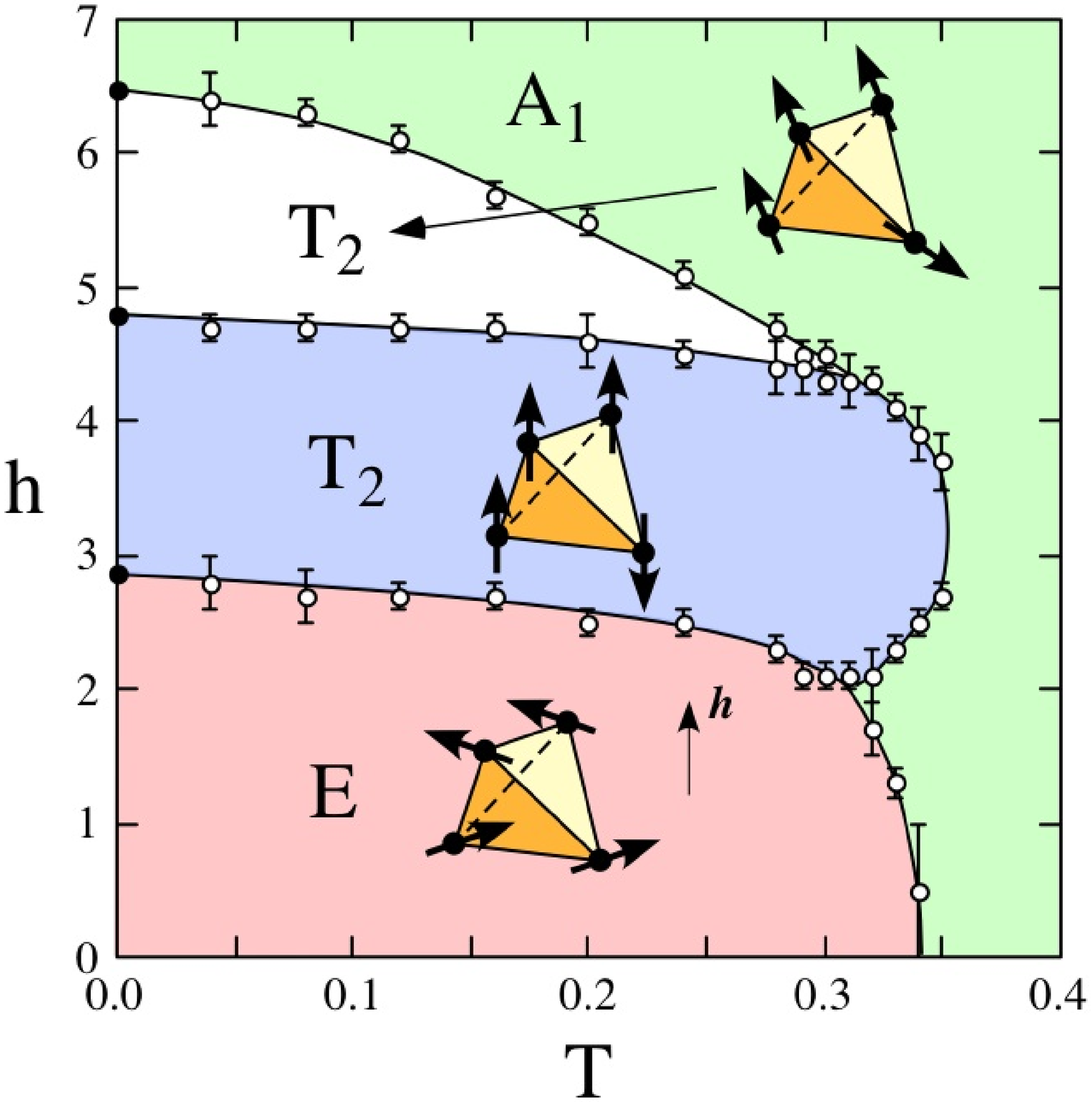}
\caption{\label{fig:theory3}Magnetic phase diagram of effective spin 
model Equation~(\ref{eqn:Hb}) for \mbox{$J_3 = -0.05$} and $b=0.1$ at finite 
temperature, taken from~\cite{motome06}.  Both $T$ and $h$ are 
measured in units of $J_1$.}
\end{minipage} 
\end{figure}

The phase diagram and magnetization process of the effective spin 
model Equation~(\ref{eqn:Hb}) at $T=0$ are shown in Figure~\ref{fig:theory2}
and Figure~\ref{fig:theory1}.  The model exhibits four dominant phases :
\begin{itemize}
\item a 2:2 coplanar canted state with tetragonal symmetry for low field 
\item a 3:1 collinear ($uuud$) half--magnetization plateau state with trigonal 
symmetry for intermediate field 
\item a 3:1 coplanar canted state with trigonal symmetry for fields approaching saturation
\item a saturated ($uuuu$) state with cubic symmetry for large magnetic field
\end{itemize}
These can be classified in terms of the two--dimensional $E$, three--dimensional $T_2$ and 
trivial $A_1$ irreducible representations (irreps) of the tetrahedral symmetry group $T_d$~\cite{penc04}.
The predicted magnetization $M(h)$ for $b \approx 0.2$ is in striking correspondence 
with experiments on Cr spinels.   

Furthermore, we can calculate the changes in bond lengths which occur in each of these phases, and do 
indeed find a giant magnetostriction on entering the half--magnetization plateau, as seen 
in experiment.   Interestingly, the (negative) sign of the magnetostriction 
seen in CdCr$_2$O$_4$ implies that the strength of AF exchange 
interactions $J_1$ {\it increases} rather than decreases with bond length, i.e.  $\alpha < 0$ in 
Equation~(\ref{eqn:H}).   This is by no means impossible in a complex 
magnetic oxide with competing exchange paths, and appears to be compatible with the 
sign of the distortion seen in the tetragonal ground state~\cite{chung05}.

The effective spin model Equation~\ref{eqn:Hb} is also amenable to low 
temperature expansion and Monte Carlo simulation 
techniques~\cite{shannon05,motome06}.   Thermal fluctuations do not significantly 
alter the phase diagram for parameters relevant to experiment, but 
impose a finite transition temperature on each of the phases described 
above.   The predicted magnetic phase diagram is shown in 
Figure~\ref{fig:theory3}.  
Once again the agreement with experiment (Figure~\ref{fig:finiteT-experiment}) is, 
for such a simple theory, striking.

The four--sublattice theory presented above is almost certainly too simple, in the sense that 
the conditions in real Cr spinels need not favour four--sublattice states.
Indeed, where known, the N\'eel and half--magnetization phases of Cr 
spinels invariably have a more complex structure.
However the present theory is none the less very successful in explaining 
their magnetization process $M(h)$.
This apparent piece of serendipity 
is in fact a profound consequence of the extreme frustration of the pyrochlore lattice.
The theory presented above accurately describes the behaviour of individual 
tetrahedra in applied magnetic field.   These tetrahedra may be 
assembled into arbitrarily complex magnetically ordered states 
without qualitative change in their thermodynamic properties (e.g. 
magnetization).   In fact it is even possible to relax the condition 
of long range magnetic order entirely, and obtain a spin liquid state 
which exhibits much the same magnetization process~\cite{shannon05}.

\section{Concluding comments}

The magnetization plateau in Cr spinels is a very recent discovery, 
and there is every reason to believe that much more can be learnt
about Cr spinels and other related systems.   
The mechanism proposed to explain 
the plateau --- a coupling to lattice degrees of freedom --- is very
general and can be expected to function in many other frustrated 
magnets and magnetic molecules.    

Many theoretical avenues also remain open, including the examination of quantum 
effects~\cite{bergman06}, the ab initio calculation of spin and 
lattice interactions~\cite{yaresko-unpub}, and the extension of 
symmetry analysis and Monte Carlo simulation to more realistic 
models~\cite{penc-unpub}.    

Half--magnetization plateaux are only one of the many 
interesting new consequences of frustration seen in 
spinel oxides.   However the possibility 
of conducting further experiments and making a quantitative comparison with 
theory means that they remain one of the most rewarding problems to study. 

\section*{Acknowledgments}
We are pleased to acknowledge helpful discussions with Masayuki 
Hagiwara, Hiroko Katori, Masaaki Matsuda, Igor Solovyev and Alexander 
Yaresko.   
NS would like to acknowledge support under the ARF program of EPSRC 
(EP/C539974/1) and the hospitality of MPI--PKS Dresden 
during the preparation of this manuscript.  
KP would like to acknowledge support under 
Hungarian OTKA Grants Nos. T049607 and K62280.  
YM would like to acknowledge support from a Grant--in--Aid for Scientific Research
(No. 16GS50219) and NAREGI from the Ministry of Education, Science, Sports, and Culture of Japan.

\end{document}